\documentstyle[aaspp4,11pt]{article}
\received{1996 September 6}
\accepted{1997 May 12}
\eqsecnum
\newcommand{\etal}{et~al.\ }
\newcommand{\eg}{e.g.,\ }

\begin{document}

\title{A PHASE-SPACE APPROACH TO COLLISIONLESS\\
STELLAR SYSTEMS USING A PARTICLE METHOD}

\author{SHUNSUKE HOZUMI}
\affil{Faculty of Education, Shiga University,\break
         2-5-1 Hiratsu, Otsu, Shiga 520, Japan; hozumi@sue.shiga-u.ac.jp}

\begin{abstract}
A particle method for reproducing the phase space of collisionless stellar
systems is described.  The key idea originates in Liouville's theorem which
states that the distribution function (DF) at time $t$ can be derived from
tracing necessary orbits back to $t=0$.  To make this procedure feasible,
a self-consistent field (SCF) method for solving Poisson's equation is
adopted to compute the orbits of arbitrary stars.  As an example, for the
violent relaxation of a uniform-density sphere, the phase-space evolution
which the current method generates is compared to that obtained with a
phase-space method for integrating the collisionless Boltzmann equation,
on the assumption of spherical symmetry.  Then, excellent agreement is
found between the two methods if an optimal basis set for the SCF technique
is chosen.  Since this reproduction method requires only the functional
form of initial DFs but needs no assumptions about symmetry of the system,
the success in reproducing the phase-space evolution implies that there
would be no need of directly solving the collisionless Boltzmann equation
in order to access phase space even for systems without any special
symmetries.  The effects of basis sets used in SCF simulations on the
reproduced phase space are also discussed.
\end{abstract}


\vfil\eject

\section{INTRODUCTION}
N-body simulations have become a powerful tool for numerical studies of
self-gravitating systems.  In fact, they have provided us with such
significant results as the bar instability in disk galaxies (\eg Hohl 1971;
Ostriker \& Peebles 1973) and the radial orbit instability in spherical
stellar systems (\eg Barnes, Goodman, \& Hut 1986).  Much of knowledge
obtained with N-body methods is essentially based on the change in, and the
evolution of, the shape of the system.  However, as far as collisionless
stellar systems like galaxies are concerned, phase-space arguments, which
take into consideration velocity space as well as configuration space, help
us understand the physics of them.  As argued by Fujiwara (1983a) in his
appendix, the core size of collapsed objects can be estimated from the
conservation of phase-space density.  In addition, such a phase-space
constraint was also applied to the explanation of why the tangential
velocity dispersion, in general, grows faster than the radial one at the
early stages of gravitational collapse for spherical systems; consequently,
the initial anisotropy in velocity dispersion was shown to be a poor
indicator of the radial orbit instability (Hozumi, Fujiwara, \& Kan-ya 1996).

Although the phase-space approach is useful for collisionless stellar
dynamics, it is difficult to access phase space numerically: such systems
are governed by the collisionless Boltzmann equation,
\begin{equation}
\frac{\partial f}{\partial t}+\mbox{\boldmath $v\cdot$}\frac{\partial f}%
{\partial \mbox{\boldmath $r$}}-\nabla \Phi \mbox{\boldmath $\cdot$}%
\frac{\partial f}{\partial \mbox{\boldmath $v$}}=0,
\eqnum{1}
\end{equation}
\noindent
and Poisson's equation,
\begin{equation}
\nabla^2 \Phi(\mbox{\boldmath $r$})=4\pi G\rho(\mbox{\boldmath $r$}),
\eqnum{2}
\end{equation}
where $f=f(\mbox{\boldmath $r$}(t),\mbox{\boldmath $v$}(t),t)$ is the
distribution function (DF), $\Phi$ the potential, $\rho$ the density,
\mbox{\boldmath $r$} the position vector, \mbox{\boldmath $v$} the velocity
vector, $t$ the time, and $G$ the gravitational constant.  Thus,
collisionless systems in general configurations must be described in
six-dimensional phase space.  It is, therefore, practically impossible to
solve equations (1) and (2) numerically due to memory limitations of
currently available computers.  One then has to approximate the system in
a simpler form.  In fact, the collisionless Boltzmann equation has been
solved numerically only for one-dimensional systems (Fujiwara 1981; Mineau,
Feix, \& Rouet 1990; White 1986), two-dimensional disk systems (Watanabe et
al. 1981; Nishida et al.\ 1981, 1984; Nishida 1986), and spherically
symmetric systems (Hoffman, Shlosman, \& Shaviv 1979; Shlosman, Hoffman,
\& Shaviv 1979; Fujiwara 1983a, 1983b; Rasio, Shapiro, \& Teukolsky 1989).
In spite of such simplification, the phase-space approach is indeed superior
to conventional N-body methods: phase-space holes became indiscernible as
time proceeded with an N-body code but were kept long-lived with a
phase-space code (Mineau \etal 1990).  Above all, Fujiwara (1983a) has
demonstrated clearly how violent relaxation, together with the subsequent
phase mixing, proceeds in phase space, and that the resulting DF in the core
is not Maxwellian but partially degenerate, which may be compared to the
prediction by Lynden-Bell (1967).

In N-body simulations as well, some efforts at representing phase space
were made to argue the kinematic nature of stellar systems.  Unfortunately,
however, only one aspect of phase-space structure was presented.  For
example, all particles were displayed on a radius versus radial-velocity plane
(H\'{e}non 1973; Min \& Choi 1989; Burkert 1990; Londrillo, Messina, \&
Stiavelli 1991); cumulative DFs in place of the fine-grained DFs were
calculated (Hernquist, Spergel, \& Heyl 1993).  Furthermore, the interaction
between a bar and a halo was analyzed on an angular-momentum versus energy
plane (Hernquist \& Weinberg 1992).  These situations arise from the fact
that the number of simulation particles is still too small to sample
six-dimensional phase space smoothly.  As a result, the conventional N-body
approach restricts our knowledge about collisionless stellar systems.
Therefore, much progress in the understanding of collisionless dynamics
should be expected if the fine-grained DF can be computed to a reliable
degree with no assumptions about the symmetry of the system.

In this paper, we give an idea to overcome that defect of the N-body
approach mentioned above.  Then, we show that the DF itself can be reproduced
even with a particle-based method.  In the reproduction process, we have no
need of assuming the symmetry of the system being studied.  In \S 2, we
describe, on the basis of Liouville's theorem and with the help of a
self-consistent field (SCF) method, how to reproduce phase space.  In \S 3,
we apply this reproduction method to the collapse of a uniform-density sphere
on the assumption of spherical symmetry, and compare the reproduced phase
space and DF it generates to those obtained with a phase-space solver.  In
\S 4, we discuss the choice of basis sets used in SCF simulations.
Conclusions are given in \S 5.

\section{REPRODUCTION METHOD OF PHASE SPACE}
If we use the Lagrangian derivative in $\mu$-space, $d/dt$, equation (1) is
rewritten as $df/dt=0$ (Binney \& Tremaine 1987).  This means that the flow
of phase elements is incompressible.  This special case of Liouville's
theorem indicates that the value of the DF at time $t$ is connected to that
of the DF at $t=0$ as
\begin{equation}
f(\mbox{\boldmath $r$}_i(t), \mbox{\boldmath $v$}_i(t), t)=%
f(\mbox{\boldmath $r$}_i(0), \mbox{\boldmath $v$}_i(0), 0),
\eqnum{3}
\end{equation}
where the subscript $i$ refers to the $i$-th phase point at time $t$.
Therefore, we can compute the DF at time $t$ by tracing the orbits of
stars from the time $t$ back to $t=0$.  Unfortunately, however, conventional
N-body techniques are incapable of following all orbits but for those of
simulation particles.  This is because usual N-body methods cannot yield
sufficiently smooth force fields with high accuracy unless a prohibitively
large number of particles are employed.

Instead, we may as well rely on an SCF method which can provide desired
force fields with a modest number of particles.  The SCF approach was first
developed by Clutton-Brock (1972, 1973), and recently, it has been revived
by Hernquist \& Ostriker (1992).  In addition, Earn \& Sellwood (1995) have
applied it to singling out the fastest growing mode of an infinitesimally
thin stellar disk using a quiet start method (Sellwood 1983).  In short,
the essence of the SCF method consists of solving Poisson's equation by
expanding the density and potential in a bi-orthogonal basis set
$(\rho_{nlm}(\mbox{\boldmath $r$}), \Phi_{nlm}(\mbox{\boldmath $r$}))$ as
\begin{equation}
\rho(\mbox{\boldmath $r$})=
\sum_{nlm} A_{nlm}(t)\rho_{nlm}(\mbox{\boldmath $r$}),
\eqnum{4}
\end{equation}
\noindent and
\begin{equation}
\Phi(\mbox{\boldmath $r$})=
\sum_{nlm} A_{nlm}(t)\Phi_{nlm}(\mbox{\boldmath $r$}),
\eqnum{5}
\end{equation}
where $n$ is the radial ``quantum" number and $l$ and $m$ are corresponding
quantities for the angular variables.  The individual harmonics $\rho_{nlm}$
and $\Phi_{nlm}$ satisfy Poisson's equation
\begin{equation}
\nabla^2\Phi_{nlm}(\mbox{\boldmath $r$})=
4\pi G\rho_{nlm}(\mbox{\boldmath $r$}).
\eqnum{6}
\end{equation}
In real simulations, the density distribution $\rho(\mbox{\boldmath $r$})$
is represented by discrete particles, and the coefficients $A_{nlm}$ at
time $t$ can be derived from equation (4) by taking advantage of the
bi-orthogonality between $\rho_{nlm}(\mbox{\boldmath $r$})$ and
$\Phi_{nlm} (\mbox{\boldmath $r$})$.  In this manner, SCF methods do not
realize perfectly smoothed force fields because they still have the effects
of particle discreteness.  Then, some kind of smoothing similar to that
caused by a softening length is included in the SCF code due to the finite
numbers of the expansion terms, though the SCF approach needs no introduction
of a softening length, as shown by equations (4) and (5).  As a result,
relaxation rates in SCF methods could be comparable to those in usual
N-body methods (Hernquist \& Barnes 1990).

However, we prefer the SCF approach to other particle methods in that we
can easily compute any orbits of stars, in addition to the fact that the
reliability of an SCF method has been demonstrated by applying it to the
collapse of spherical stellar systems (Hozumi \& Hernquist 1995, hereafter
HH).  Once the coefficients $A_{nlm}$ are found, the acceleration
$\mbox{\boldmath $a$}(\mbox{\boldmath $r$})$ can be obtained through
equation (5) as
\begin{equation}
\mbox{\boldmath $a$}(\mbox{\boldmath $r$})=-\sum_{nlm} A_{nlm}(t)%
\nabla\Phi_{nlm}(\mbox{\boldmath $r$}),
\eqnum{7}
\end{equation}
where $\nabla\Phi_{nlm}(\mbox{\boldmath $r$})$ is, of course, calculated
analytically in advance when a basis set is given.  We can see from
equation (7) that the SCF technique is capable of computing force fields at
any times if we save the coefficients $A_{nlm}$ at the corresponding times.
When simulating collisionless systems, one can advance the coordinates of
particles by $\Delta t$ in time with a suitable integration scheme, so that
one can use the computed orbits to find the evolving density distribution,
recompute the potential and iterate.  In this way, we first run an SCF
simulation, storing the coefficients $A_{nlm}$ at each time step.  Next, we
trace the necessary orbits of stars back to $t=0$ using the coefficients
$A_{nlm}$.  Consequently, the DF at time $t$ can be reproduced in terms of
equation (3).  Since the computation of DFs with the current method includes
no particular numerical diffusion, DFs like those reproduced here are, in this
sense, regarded as the fine-grained DFs.

\section{REPRODUCED PHASE SPACE}
\subsection{Model and Numerical Procedure}
We choose the same model as was used in HH to apply the reproduction method
of phase space explained in \S 2.  The model consists of a uniform-density
sphere having a Maxwellian velocity distribution with the initial virial
ratio of 1/2.  Then, the DF, $f$, is written as
\begin{equation}
f=\left(\frac{3M}{4\pi R_0^3}\right)\left(\frac{1}{2\pi\sigma_0^2}
\right)^{3/2}
\exp [-\frac{(u^2+j^2/r^2)}{2\sigma_0^2}],
\eqnum{8}
\end{equation}
where $M$ is the total mass, $R_0$ the radius, and $\sigma_0$ the velocity
dispersion with $r$, $u$, and $j$ being the radius, radial velocity, and
angular momentum, respectively.

This model has been shown by HH that the relaxed density and velocity
dispersion profiles obtained from the SCF simulation are in excellent
agreement with those from a collisionless Boltzmann simulation on the
assumption of spherical symmetry.  Then, we further examine how well the
evolution in phase space can be reproduced with recourse to the SCF
technique on the same assumption.  In the SCF code, we can accomplish
spherical symmetry by retaining only $l=m=0$ terms in the expansions of
the angular variables.

The simulation procedure for the SCF technique is the same as that adopted
in HH.  The units of the gravitational constant and mass are such that $G=1$
and $M=1$, respectively.  We use $R_0=2$, and choose the length scale of the
basis functions to be unity.  In these units, the time required for the
collapse of an exactly ``cold" sphere, $T_{\rm c}$, turns out to be $\pi$.
Then, the time step is chosen to be $\Delta t=0.05$, half as small as in HH
to compute the orbits more accurately.  The maximum number of the radial
expansion coefficients, $n_{\rm max}$, is taken to be 32.  We employ
$N=100,000$ particles of equal mass.  The equations of motion are integrated
in Cartesian coordinates using a time-centered leapfrog algorithm (\eg
Press \etal 1986).  We use the same integration algorithm for tracing the
orbits of stars to compute the values of the DF.  We employ two types of
basis set: one is that constructed by Clutton-Brock (1973, hereafter the CB
basis set) which is derived from the Plummer model for spherical stellar
systems (Plummer 1911; Binney \& Tremaine 1987), and the other is that
developed by Hernquist and Ostriker (1992, hereafter the HO basis set),
being based on the model for spheroidal galaxies proposed by Hernquist (1990).

For the purpose of comparison as was done in HH, we again solve the same
problem using a phase-space solver developed by Fujiwara (1983a) with the
mesh points $(N_r, N_u, N_j)=(200, 200, 50)$, where $N_r$, $N_u$, and $N_j$
are the numbers of mesh points along radius, radial velocity, and angular
momentum, respectively.  The collisionless Boltzmann equation is integrated
using a splitting scheme (Cheng \& Knorr 1976).  To reduce the numerical
diffusion generated from the repeated interpolation required by the splitting
scheme, we trace the orbits of the stars on the grid points backward to $t=0$
at times when the phase-space representation is needed.  This line of
numerical sophistication was first devised by Rasio \etal (1989).  The other
parameters are $R_{\rm min}=0.01$, $R_{\rm max}=15.9$, $U_{\rm max}=2.1$, and
$J_{\max}=1.7$ in Fujiwara's (1983a) notation.  The time step used is
$\Delta t=0.05$.

\subsection{Evolution in Phase Space}
For the construction of phase space, we compute all the orbits of those stars
which correspond to the grid points of the collisionless Boltzmann simulation,
from time $t$ backward to $t=0$.  In so doing, two angular momenta such as
$j=0.00068$ and $j=0.329$ are employed.  Thus, we carry out the orbit
computation for $N_r \times N_u=40,000$ points on each $j$-plane.

\begin{figure}
\figurenum{1}
\plotone{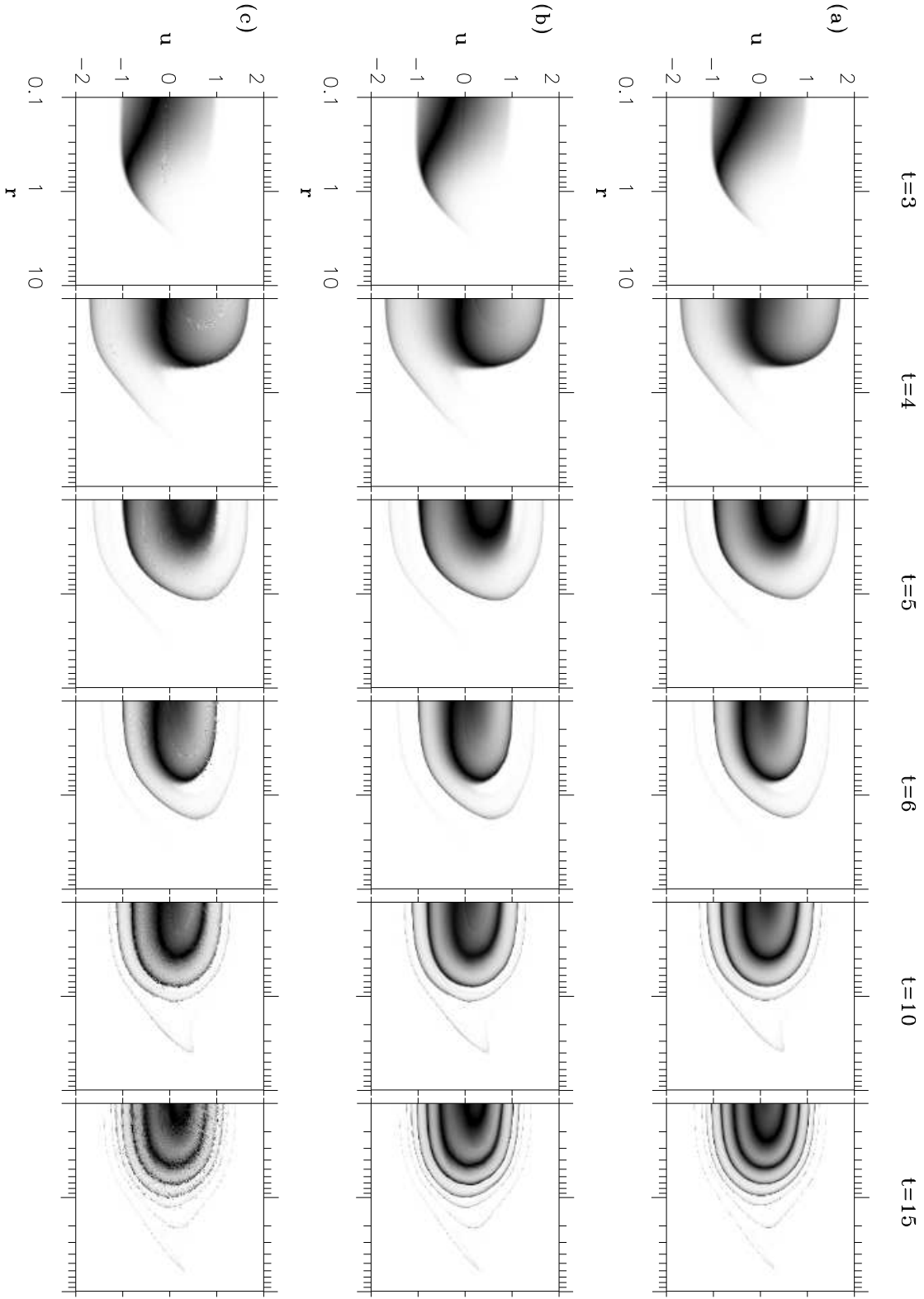}
\caption{Phase-space evolution for the collapse of a uniform-density
sphere on a $j=0.00068$ plane derived from the collisionless Boltzmann
simulation (a), reproduced with the CB basis set (b), and with the HO basis
set (c).  The initial virial ratio is 1/2.  The ordinate is the radial
velocity, and the abscissa is the radius in a logarithmic scale.}
\end{figure}

\begin{figure}
\figurenum{2}
\plotone{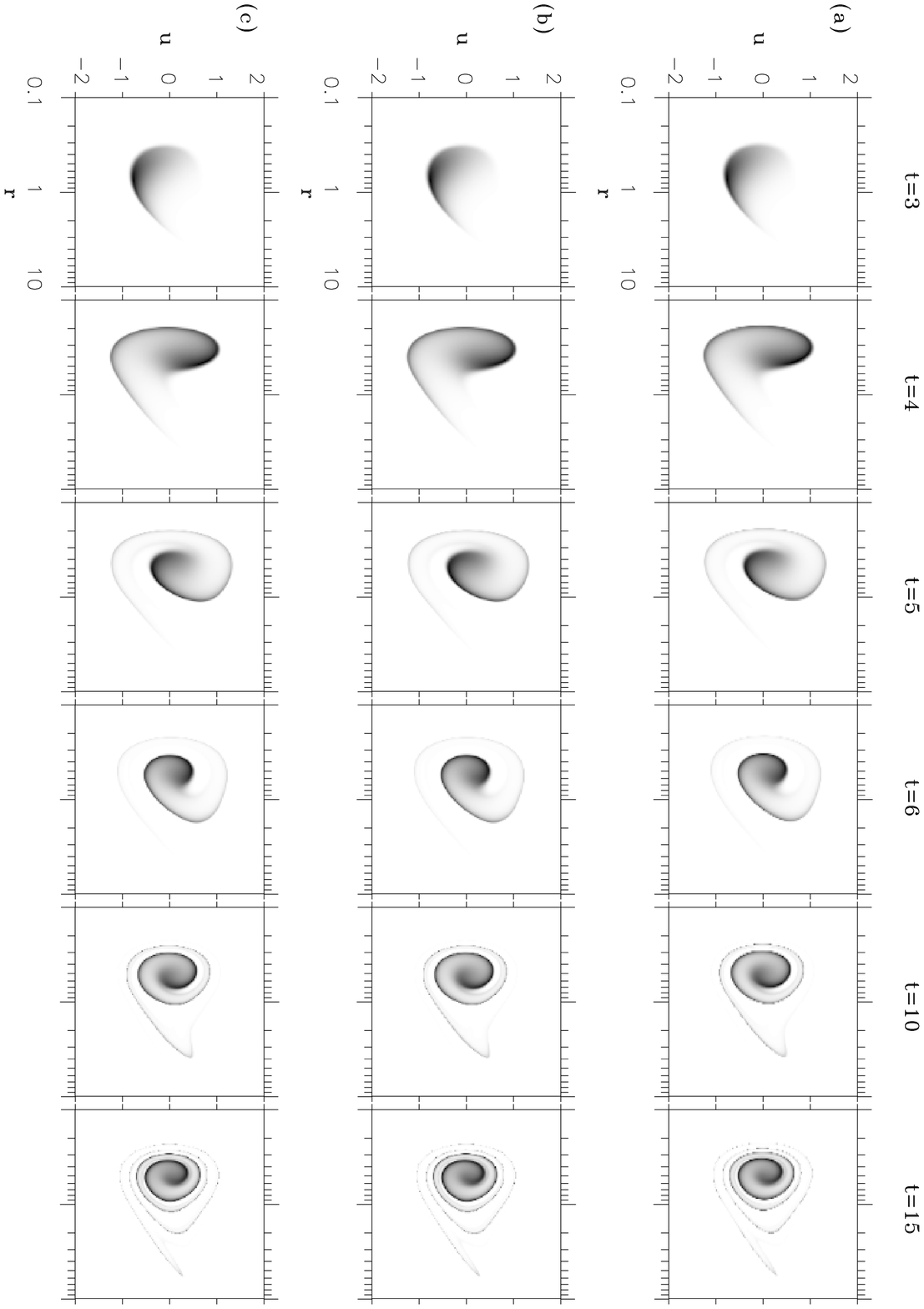}
\caption{Same as in Fig. 1, but on a $j=0.329$ plane.}
\end{figure}

We show in Figures 1 and 2 the phase-space evolution reproduced with the CB
and HO basis sets, along with that derived from the collisionless Boltzmann
simulation.  The detailed description of the collapse, beginning from uniform
contraction through violent relaxation to phase mixing, is given by Fujiwara
(1983a).  It can be noticed that the evolution is essentially the same
between the two SCF simulations, except for a ragged structure appearing on
a $j=0.00068$ plane for the HO basis set.  The origin of this unfavorable
structure is discussed in the next section.  On the other hand, a rough
comparison in Figure 1 reveals that the phase-space evolution proceeds
similarly in an essential sense between the phase-space solver and the SCF
method with the CB basis set.  However, a closer look reveals the difference
which started to become noticeable at $t=6$, though the subsequent evolution
did not deviate substantially between the two methods.  In contrast to the
evolution for the small value of the angular momentum, Figure 2 which
represents phase space on a somewhat large angular-momentum plane
demonstrates that there is no practical difference in phase-space
evolution between the two methods, even when the HO basis set was used in
the SCF simulation.  In spite of the existing difference found in Figure 1,
these figures manifest that the method of reproducing phase space proposed
in this paper is competitive with that of directly integrating the
collisionless Boltzmann equation as far as spherically symmetric systems
are concerned.

Figure 1 indicates that the choice of basis sets affects the degree of the
reproduction quality on small angular-momentum planes, that is, for the
regions dominated by the stars passing close to the center.  Clearly, the
CB basis set is more suitable for reproducing phase space on every
angular-momentum plane than the HO basis set.  Nevertheless, we can
understand from the same figure that the ragged structure resulting from the
HO basis set will have no serious effects on the low order moments of the DF
such as the density and velocity dispersions when such moments are evaluated
directly from N particle distributions, because the velocity spread does not
differ considerably between the two basis sets even on such small
angular-momentum planes.  In fact, this claim has already been demonstrated
by HH who showed the excellent agreement in density and velocity dispersion
profiles calculated from simulation particles though they used cooled
Plummer models.

\subsection{Distribution Function at a Relaxed State}
The reproduction of the DF is another important aspect of the phase-space
approach to collisionless dynamics.  As argued by Fujiwara (1983a),
incompleteness of violent relaxation leads to a degenerate DF in the
core.  Though Fujiwara (1983a) assumed spherical symmetry, May \& van
Albada (1984), using their expansion code, showed that the phase-space
density in the core did not decrease substantially from its initial value
for the three-dimensional collapse.  Unfortunately, they estimated the
phase-space density only in the mean sense like
$\rho_{\rm c}/\sigma_{\rm c}^3$, where $\rho_{\rm c}$ and $\sigma_{\rm c}$
are the density and velocity dispersion in the core, respectively.  We then
demonstrate that our reproduction method can yield the DF itself.

\begin{figure}
\figurenum{3}
\plotone{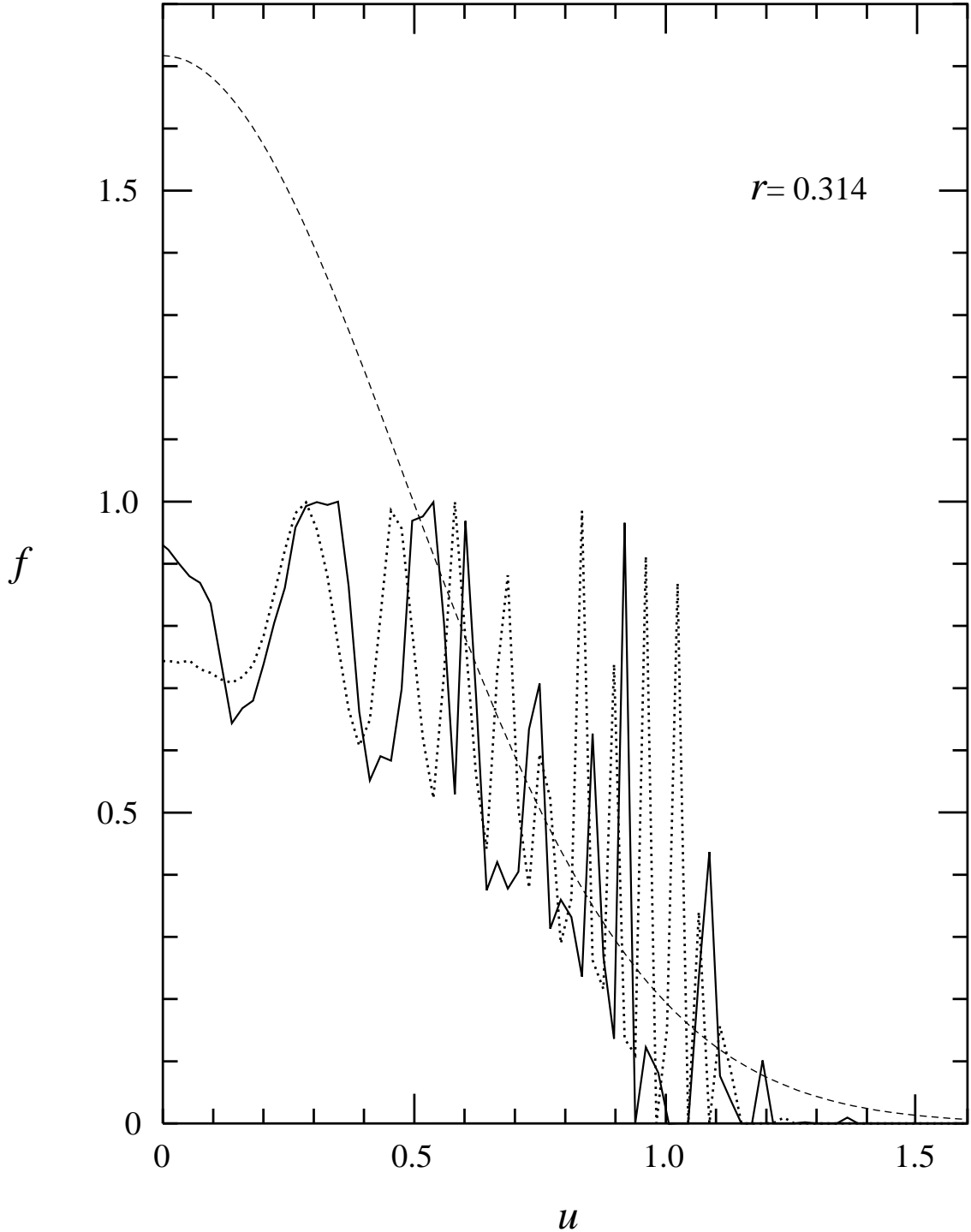}
\caption{Distribution functions $f=f(r, u, j=0)$ in the core
($r=0.314$) for a relaxed state ($t=50$) of the uniform-density sphere,
normalized by the initial maximum value of
$(3M/4\pi R_0^3)(2\pi \sigma_0^2)^{-3/2}$, as a function of $u$.  The solid
line shows the DF reproduced with the CB basis set and the dotted line the
DF derived from the collisionless Boltzmann simulation.  The dashed curve
represents a Maxwellian distribution function constructed with the density,
$\rho(r)$, and the velocity dispersion, $\sigma^2$, at $t=50$ such that
$f=\rho(r)(2\pi \sigma^2)^{-3/2}{\rm exp}(-u^2/2\sigma^2)$ with
$\sigma^2=(\langle u^2\rangle +\langle j^2/r^2\rangle)/3$, normalized by
$(3M/4\pi R_0^3)(2\pi \sigma_0^2)^{-3/2}$.}
\end{figure}

In Figure 3, $f(r, u, 0)$ in the core ($r=0.314$) at a relaxed state
($t=50$) is presented for the SCF simulation with the CB basis set and
the collisionless Boltzmann simulation, which corresponds to Figure 6 of
Fujiwara (1983a).  We show only the $u>0$ part of the DF because of its
practical symmetry about $u$.  We further mention that $f(r, 0, j)$ in
the core is very similar to $f(r, u, 0)$, as shown by Fujiwara (1983a).
Then, it can be noticed that the DF reproduced with our method is in good
agreement with that derived from the phase-space method.  In addition,
the relaxed core is evidently degenerate because the values of the DF
normalized by the maximum value of the initial DF are close to the unity
for small values of $u$.  To make clear the degeneracy, we have added
a Maxwellian curve constructed from the density and velocity dispersions
at $t=50$.  This finding has already been pointed out by Fujiwara (1983a)
who gave a possible mechanism of this degeneracy.  The fluctuating behavior
of the DFs reflects smaller and smaller structures generated from phase
mixing with time, as stated by Lynden-Bell (1967).  Figure 1 is also helpful
to understand this behavior.

\section{DISCUSSION}
We argue the origin of the ragged structure appearing in the reproduced phase
space with the HO basis set.  Its lowest order member of the potential basis
functions is reduced to the Hernquist model (Hernquist 1990):
\begin{equation}
\Phi_{000}^{\rm HO}(r)=-\frac{1}{1+r},
\eqnum{9}
\end{equation}
in a dimensionless system of units.  This potential generates the nonzero
radial force, $f_r$, at the center: $f_r(0)=-1$.  In reality, the force
should converge to zero as $r\rightarrow 0$.  Thus, the stars suffer
fictitious forces when passing close to the center, so that their orbits are
greatly displaced from the true orbits.  This effect then emerges, as shown
in Figure 1(c), as a ragged structure on small angular-momentum planes where
stars orbiting close to the center are dominant.  Of course, raggedness can
be mitigated by choosing a smaller time step.  In fact, with $\Delta t=0.0125$
and the other parameters intact, we recovered smoother phase plots than those
shown in Figure 1(c) (not presented here), but we have found that there still
remains raggedness.  As a result, for small values of the angular momentum,
we need much more computational cost to attain a smooth representation of
phase space with the HO basis set than with the CB basis set.  However, the
nonvanishing force at the center no longer affects the orbits which avoid
approaching the center closely.  Therefore, we can find no unusual feature
in the reproduced phase space of Figure 2(c) even with the HO basis set.  On
the other hand, since the zeroth order term of the potential basis functions
for the CB basis set is the Plummer model (Plummer 1911; Binney \& Tremaine
1987) given by
\begin{equation}
\Phi_{000}^{\rm CB}(r)=-\frac{1}{(1+r^2)^{1/2}},
\eqnum{10}
\end{equation}
the resulting force converges to zero as $r\rightarrow 0$.  Consequently,
the reproduced phase space with the CB basis set is a smooth distribution
of phase particles.

We should point out that the phase-space solver used here also has a
difficulty in following phase-space evolution for small values of the
angular momentum.  Since we set up a reflection wall at $R_{\rm min}$,
the stars which have reached the region within $R_{\rm min}$ from outside
are forced to be moved outside of $R_{\rm min}$ almost instantaneously with
the sign of the radial velocity reversed.  This means that the phase-space
evolution on small angular-momentum planes proceeds faster than real,
each time the stars traverse the reflection wall from outside, by
approximately the crossing time of $2R_{\rm min}/U$, where $U$ is the
typical radial velocity at $R_{\rm min}$.  Indeed, when we adopted
$R_{\rm min}=0.1$, we found that phase particles in the collisionless
Boltzmann simulation were wound up many more times than those in the SCF
simulations on a $j=0.00068$ plane because of the effect just mentioned.
However, the phase-space evolution with $R_{\rm min}=0.005$ no longer
showed any difference from that with $R_{\rm min}=0.01$.

\begin{figure}
\figurenum{4}
\plotone{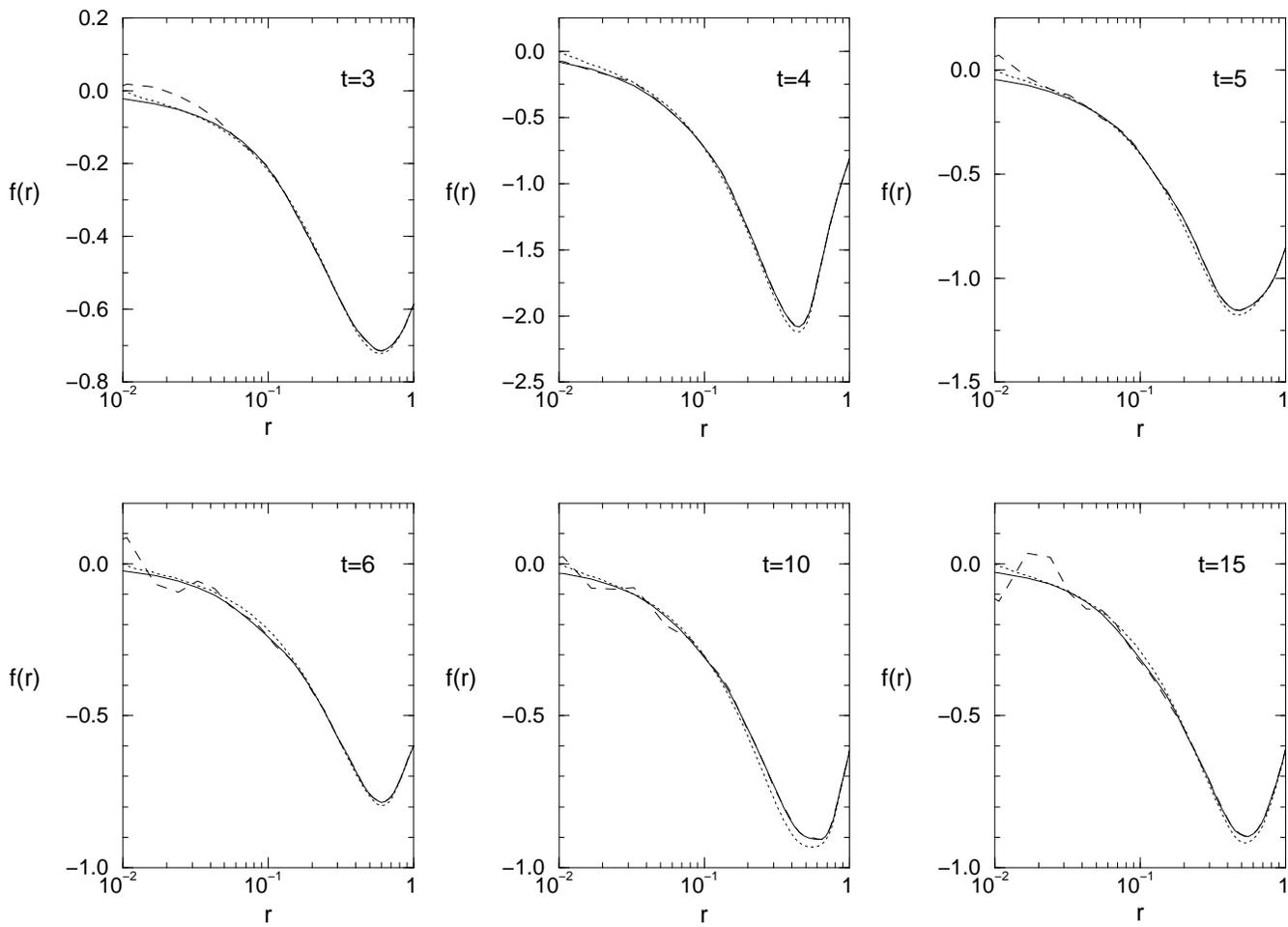}
\caption{Time evolution of radial forces, $f(r)$, for 
the uniform-density sphere as a function of radius.  The solid lines are
obtained with the CB basis set, the dashed lines with the HO basis set,
and the dotted lines from the collisionless Boltzmann simulation.}
\end{figure}

On the other hand, SCF methods inherently have relaxation effects similar
to those of the softening length as pointed out by Hernquist \& Barnes
(1990), so that the phase-space evolution obtained with the SCF method would
not necessarily reproduce the true evolution.  In this way, it is difficult
to evaluate the numerical accuracy of phase-space evolution for the
collisionless Boltzmann and SCF simulations.  One measure for the accuracy
may be to compare the evolving radial forces between the two methods.  Then,
we present in Figure 4 the radial forces at various times.  We can notice
from this figure that the forces expanded with the CB basis set are in
good agreement with those from the collisionless Boltzmann simulation
though there is a slight difference around the minimum values, which could
come from the smoothing effects caused by the finite expansion terms in
the SCF code.  As compared to the CB basis set, the HO basis set gives a
poor representation of the radial forces at small radii.  In particular,
within $r \sim 0.1$, the forces expanded with the HO basis set deviate
appreciably from those with the CB basis set and become even positive around
$R_{\rm min}=0.01$ except at $t=4$.  Thus, we may understand again why the
ragged structure appeared on a small angular-momentum plane when we used the
HO basis set.  Owning to the reflection wall within which there is no mass,
the forces at $R_{\rm min}$ become exactly zero for the collisionless
Boltzmann simulation.  In this sense, the forces obtained with the CB basis
set are more faithful to the real ones than those with the phase-space solver.

In addition to the phase-space evolution, the reproduced DF with the SCF
technique is in good agreement with that using the phase-space solver, as
shown in Figure 3.  Besides, the reproduced DF is considered the fine-grained
DF because our computation method of the DF suffers no coarse-graining.
Since our reproduction method is easily applied to stellar systems
with no special symmetries, we will be able to precisely study the degree
of the degeneracy of the relaxed cores for three-dimensional collapses.
In order to extend our method to general configurations, we again only save
the expansion coefficients, $A_{nlm}$, at each time step to trace the orbits
of stars backward to $t=0$.  Of course, the quantum numbers, $l$ and $m$,
have nonzero terms in this case.  Although the extension of our method is
quite straightforward, the development of smaller and smaller structures in
phase space with time will force us to sample a large number of orbits for
the accurate reproduction of the DF.

Since the success of the reproduction method proposed in this paper depends
on the computation accuracy of individual orbits, a successful representation
of the phase-space evolution, as demonstrated here, implies that the SCF
technique enables us to compute the orbits of stars very close to the true
orbits except for those passing near the center of the system.  Therefore,
the classification of orbit families like that done by van Albada (1987) can
be made more reliably with the SCF method than with conventional N-body
methods.  In particular, SCF algorithms are well-suited for parallelized
computer architectures (Hernquist \& Ostriker 1992; Hernquist, Sigurdsson,
\& Bryan 1995), so that they can suppress discreteness noise by employing
$N\gtrsim 10^6$.  These algorithms will thus make feasible the orbit
classification with sufficiently high accuracy.  Furthermore, an algorithm,
originally noticed by Merritt (1996) and recently realized by Weinberg (1996),
for optimally choosing the maximum order term in the expansions such as
$n_{\rm max}$ will be useful to give closer force fields to the exact ones.

\section{CONCLUSIONS}
We have demonstrated that the phase-space evolution for the collapse of a
uniform-density sphere can be reproduced even using a particle method on the
basis of Liouville's theorem and with recourse to the SCF approach though
spherical symmetry is assumed.  Fortunately, our reproduction method is
applicable to the systems with no special symmetries in contrast to the
phase-space method whose application is restricted to the systems with some
kind of symmetry due to memory limitations of presently available computers.
The application of the present method to general aspherical systems is
straightforward and can be realized only by saving the expansion coefficients,
$A_{nlm}$, at each time step, though the computational cost increases in an
SCF run and the orbit tracing because of the additional terms in the angular
expansions.  Therefore, we could no longer need to depend on collisionless
Boltzmann simulations in order to access phase space, at least on such
problems as the stability and collapse of the stellar systems having no
symmetry.

For a good construction of phase space, we should choose such a basis
set in the SCF code that the expanded forces converge to zero as $r
\rightarrow 0$.  If we use a basis set which gives rise to nonzero forces as
$r \rightarrow 0$, the reproduced phase space will show a ragged structure
for small values of the angular momentum.  In this respect, the CB basis set
which has no unusual characters in force is more suitable to the reproduction
of phase space than the HO basis set whose expanded forces become nonzero at
the center.  However, the difference in the reproduced phase space resulting
from the two basis sets disappears if somewhat large values of the angular
momentum are chosen to plot phase particles.

We can also construct the DF itself with the SCF technique, which is found
to be in good agreement with that obtained from the collisionless Boltzmann
simulation.  The reproduced DF is considered the fine-grained DF because of
no coarse-graining included with our method.  In addition, the present method
for constructing the DF can be used to examine the degree of the degeneracy
of the relaxed core for aspherical collapses of stellar systems.

The success of our reproduction method means that the individual orbits of
stars are computed accurately.  Thus, SCF algorithms can apply to the
classification of orbit families to understand the kinematic properties of
galaxies.  Since the SCF method can employ a sufficiently large number of
particles owning to its perfectly scalable nature, the orbit classification
will be made with high accuracy by greatly reducing discreteness noise.

\acknowledgments The author would like to thank Dr.\ T.\ Fujiwara for
stimulating discussions on the reproduction of phase space.  Thanks are also
due to Dr.\ L.\ Hernquist for providing the author with his original SCF
code and valuable comments on the manuscript.  Professors S.\ Kato and
S.\ Inagaki are acknowledged for their critical reading of the manuscript.
Concerning the preparation of Figures 1 and 2, Aki Takeda is greatly
acknowledged.  This work was supported in part by a Grant-in-Aid for
Scientific Research from the Ministry of Education, Science, Sports, and
Culture of Japan (08740169).

\end{document}